\documentclass[10pt]{article}

\usepackage{graphics}
\usepackage{amscd}
\usepackage{amsmath}
\usepackage{amsfonts}
\usepackage{amssymb}
\usepackage{cite,color}

\newcommand{\bF}{ {\mathbb F}}

\newcommand{\N}{ {\mathbb N}}

\newcommand{\cI} { \mathcal{I} }
\newcommand{\2} {$2$-to-$1$}
\newcommand{\EOP} { \hfill $\Box$ }

\newtheorem{thm}{Theorem}[section]

\newtheorem{lem}[thm]{Lemma}
\newtheorem{prop}[thm]{Proposition}

\newtheorem{rem}[thm]{Remark}
\newtheorem{cor}[thm]{Corollary}

\newtheorem{defi}[thm]{Definition}

\newenvironment{pf}[1][\it Proof]{\textbf{#1.}}{\ \rule{0.5em}{0.5em}}

\begin{document}

\title{Two-to-one mappings and involutions without fixed points over~$\bF_{2^n}$}


\author{ Mu Yuan$^1$, Dabin Zheng$^1$\footnote{Corresponding author, yuanmu847566@outlook.com (M.\ Yuan), dzheng@hubu.edu.cn(D.\ Zheng), ypwang@aliyun.com(Y.\ Wang).
This work was partially supported by the National Natural Science Foundation of China under Grant Number 11971156. }\,\,and\, Yan-Ping Wang$^2$ }

\date{\small 1. Hubei Key Laboratory of Applied Mathematics, \\
Faculty of Mathematics and Statistics, Hubei University, Wuhan 430062, China \\
2. College of Mathematics and Statistics, Northwest Normal University, Lanzhou 730070, China
}

\maketitle

\leftskip 1.0in
\rightskip 1.0in

\noindent {\bf Abstract.}
In this paper, two-to-one mappings and involutions without any fixed point on finite fields of even characteristic are investigated. First, we characterize a closed relationship between them by implicit functions and develop an AGW-like criterion for \2 mappings. Using this criterion, some new constructions of \2 mappings are proposed and eight classes of \2 mappings of the form $(x^{2^k}+x+\delta)^{s}+cx$ are obtained. Finally, a number of classes of involutions without any fixed point are derived from the known \2 mappings by the relation between them.

\vskip6pt

\vskip 35pt

\leftskip 0.0in
\rightskip 0.0in

\section{Introduction}

Let $q$ be a power of a prime $p$ and $\bF_{q}$ be the finite field with $q$ elements. A polynomial $f(x)\in \bF_{q}[x]$ is called a permutation polynomial if $f: c\mapsto f(c)$ from $\bF_q$ to itself is a bijection. In particular, $f$ is called an {\it involution} if the compositional inverse of $f$ is itself. If involutions are used as part of a block cipher, they are very suitable for use in devices with  limited resources since the implementation of the inverse does not require additional resources. Involutions have been used in block ciphers such as AES~\cite{AES}, Khazad, Anubis~\cite{Biryukov2003}
and PRINCE~\cite{Borghoff2012}. Furthermore, to resist reflection cryptanalysis, the involutions used in PRINCE-like ciphers should not have fixed points~\cite{BouraCanteautKnudsenLeander2017,SoleimanyBlondeauYu2015}. The reference~\cite{CharpinMesnagerSarkar2016} first provided a systematic study of involutions over finite field of even characteristic and recent results on involutions can be found in~\cite{CharpinMesnagerSarkar2015,CharpinMesnagerSarkar2016,FuFeng2018,NiuLiQuWang2020,Rubio2017,ZhengYuanLiHuZeng2019}.
A polynomial $f(x)\in \bF_{q}[x]$ is said to be a two-to-one (\2) mapping on $\bF_{q}$ if the equation $f(x)=a$ has either two or no solutions in $\bF_{q}$ for any $a \in \bF_{q}$. Two-to-one mappings over finite fields are used in design special important primitives in symmetric cryptography such as APN functions~\cite{Pott2016,Villa2019}, bent functions~\cite{CarletMesnager2011}, planar functions~\cite{ChenPolhill2011,WengZeng2012} and good linear codes~\cite{LiLiHellesethQu2020} in coding theory. For example, Carlet and Mesnager constructed new bent Boolean functions in bivariate representation with \2 mappings over finite fields of even characteristic~\cite{CarletMesnager2011}. In addition, the hyperoval in finite geometry can be defined by o-polynomials, which are closed related to 2-to-1 mappings. As far as we know, Charpin and Kyureghyan in~\cite{CK2009,K} first obtained some \2 mappings of the special forms in their study of permutation polynomials by using linear translators. Very recently, Mesnager and Qu in~\cite{MesnagerQu2019} first systematically studied \2 mappings over finite fields as a specific goal, then Li et al.~\cite{LiMesnagerQu2019} further constructed some \2 trinomials and quadrinomials over finite fields of even characteristic. Almost all \2 mappings over $\bF_{2^n}$ in~\cite{MesnagerQu2019,LiMesnagerQu2019} exist only when $n$ is odd.

In this paper, we first characterize a relationship between \2 mappings and involutions over finite fields and show that the involutions derived from \2 mappings have no
fixed points. Then, along the line of work in~\cite{MesnagerQu2019}, we propose another AGW-like criterion for constructing \2 mappings over $\bF_{2^n}$. By application
of this criterion, some general constructions of \2 mappings over $\bF_{2^n}$ are obtained. In particular, we investigate the \2 property of polynomials over $\bF_{2^{2m}}$ with the following form
\begin{equation}\label{eq:main}
(x^{2^k}+x+\delta)^s+cx,
\end{equation}
and obtain many new \2 mappings of the form~(\ref{eq:main}) and corresponding involutions derived from them over $\bF_{2^{2m}}$ by using the proposed relation and the multivariate method introduced by Dobbertin~\cite{Dobbertin2002}. Moreover, some involutions over $\bF_{2^{2m+1}}$ are also derived from known \2 mappings over this finite field.

The remainder of this paper is organized as follows. In Section~\ref{sec:Preliminaries}, we introduce some basic concepts and related results. In Section~\ref{sec:relation}, we characterize a relationship between \2 mappings and involutions without any fixed point. In Section~\ref{sec:AGW-like criterion}, we provide another AGW-like criterion for constructing \2 mappings over finite fields. In Section~\ref{sec:\2 mappings}, many \2 mappings with the form (\ref{eq:main}) over $\bF_{2^{2m}}$ are obtained and the corresponding involutions are derived from them. In Section~\ref{sec:involution for odd n}, many involutions without any fixed point over $\bF_{2^{2m+1}}$
are derived from known \2 mappings over this finite field. Finally, Section~\ref{sec:final} concludes this paper.


\section{Preliminaries}\label{sec:Preliminaries}

Let $q$ be a power of $2$, $\bF_{q}$ the finite field with $q$ elements and $\bF_q^*$ its multiplicative group. Let $\N$ denote the set of all positive integers.
For $d \in \N$ with $d \,|\, (q-1)$, let $\mu_d$ denote the subgroup of $\bF_{q}^*$ with order $d$ and $\mu_{d}^*=\mu_{d} \setminus \{1\}$.
For $k, \ell \in \N$ with $\ell\, | \,k$, the trace function ${\rm Tr}^k_{\ell}(\cdot)$ is defined as
\begin{equation*}
  {\rm Tr}^k_{\ell}(x)=\sum_{i=0}^{k/\ell-1}x^{2^{\ell\cdot i}}.
\end{equation*}
In particular, ${\rm Tr}^k_1(x)$ is called the absolute trace function defined on $\bF_{2^k}$, and it is denoted by ${\rm Tr}_{k}(x)$ for short.
A positive integer $s$ is called a {\it Niho exponent} with respect to the finite field $\bF_{2^{2m}}$ if $s\equiv 2^j \,\, ({\rm mod }\,\, 2^m-1)$ for some
$j \in \N$. Now, we give the definition of 2-to-1 mappings over any finite set.

\begin{defi}\cite{MesnagerQu2019}
Let $A$ and $B$ be two finite sets, and let $f$ be a mapping from $A$ to $B$. Then $f$ is called a 2-to-1 mapping if one of the following two cases holds:
\begin{enumerate}
\item[{\rm (1)}] If $\sharp A$ is even, then for any $b \in B$, it has either $2$ or $0$ preimages of $f$;
\item[{\rm (2)}] If $\sharp A$ is odd, then for all but one $b \in B$, it has either $2$ or $0$ preimages of $f$, and the exception element has exactly one preimage.
\end{enumerate}
\end{defi}
Next, we recall the resultant of two polynomials over fields. Let $f(x)=\sum_{i=0}^n a_{n-i} x^i$ and $g(x)=\sum_{i=0}^m b_{n-i} x^i$ be two polynomials of positive degree over $\bF_q$,  where $a_0\neq 0$ and $b_0\neq 0$.
The Sylvester resultant of $f(x)$ and $g(x)$, denoted by $Res(f,g,x)$, is the following determinant:
\begin{eqnarray*}
Res(f, g, x)=
\begin{array}{|cccccccc|}
   a_{0} &    a_{1}    & \cdots &   a_{n}  & 0  &    &\cdots & 0   \\
    0  &  a_{0} &    a_{1}    & \cdots &   a_{n}  & 0  & \cdots & 0 \\
    \vdots  &    &        &      &     &      &          &  \vdots    \\
    0  &  \cdots &   0   &  a_{0} &    a_{1}  &     &  \cdots &   a_{n} \\
    b_{0} &    b_{1}    & \cdots &      & b_{m}  & 0  & \cdots &  0     \\
      0   &    b_{0} &   b_{1}  & \cdots &    &  b_{m}  & \cdots &  0 \\
    \vdots &          &      &     &      &     &    &\vdots       \\
    0  &   \cdots   & 0 &  b_{0} &    b_{1} &    & \cdots &  b_{m}  \\
\end{array}
& \begin{array}{l}
\left.\rule{0mm}{10mm}\right\}$m$~\text{rows}\\
\\\left.\rule{0mm}{10mm}\right\}$n$~\text{rows}
\end{array}\\[0pt]
\end{eqnarray*}

There are two polynomials $A, B \in \bF_q[x]$ such that $Af+Bg=Res(f,g,x)$, and so $Res(f,g,x)=0$ if and only if $f(x)$ and $g(x)$ have a common divisor of positive degree.
For two polynomials $F(x,y),\, G(x,y)\in \mathbb{F}_q[x,y]$ of positive degree in $y$, the resultant $Res(F,G,y)$ with respect to $y$ is a polynomial in $x$, and is in the elimination ideal
$\langle F, G\rangle \cap\bF_q[x]$. So, $Res(F,G,y)$ vanishes at any common solution of $F=G=0$. For more information on resultants, the reader can refer to~\cite{Cox2007}.

For later usage we recall some results on the number of the solutions of quadratic and cubic equations over $\bF_{2^n}$.

\begin{lem}\label{lem:square}\cite{LidlNiederreiter1983}
Let $a, b \in \bF_{2^n}$ and $a \neq 0$. Then the quadratic equation $x^2+ax+b=0$ has solutions in $\bF_{2^n}$ if and only if ${\rm Tr}_{n}(\frac{b}{a^2})=0$.
\end{lem}

\begin{lem}\label{lem:cubic}\cite{BerlekampRumseyOlomon}
Let $a,b \in \bF_{2^n}$ and $b \neq 0$. Then the cubic equation $x^3+ax+b=0$ has a unique solution in $\bF_{2^n}$ if and only if ${\rm Tr}_{n}(\frac{a^3}{b^2}+1)\neq0$.
\end{lem}

\begin{lem}\label{lem:quartic}\cite{MesnagerQu2019}
Let $f(x)=x^4+a_3x^3+a_2x^2+a_1x \in \bF_{2^n}[x]$. Then $f(x)$ is \2 over $\bF_{2^n}$ if and only if one of the following holds:
\begin{itemize}
\item $a_3=a_1=0$, $a_2\neq 0$;
\item $a_3=0$, $a_1\neq 0$ and ${\rm Tr}_{n}\left(a_2^3/a_1^2+1 \right)\neq 0$;
\item $n$ is odd, $a_3 \neq 0$ and $a_2^2=a_1a_3$.
\end{itemize}
\end{lem}

\section{A relation between \2 mappings and involutions}\label{sec:relation}

In this section, a closed relationship between \2 mappings and involutions over finite fields is investigated.  This relation is useful for us to obtain
involutions without fixed points from 2-to-1 mappings over finite fields.  We also study the relation between two involutions derived from two \2 mappings
which are differed by a bijection.

\begin{thm}\label{thm:main}
Let $D$ be a subset of $\bF_q$ with even cardinality and $\mathcal{D}=\{(x,y) \in D^2| \,\,\,\,  x \neq y\}$. Let $f$ be a \2 mapping on $D$. Then the equation $G(x,y)=f(x)-f(y)=0$ on $\mathcal{D}$ determines an involution $ y=\cI_f(x)$ on $D$. Furthermore, $y=\cI_f(x)$ has no fixed point on $D$.
\end{thm}

\pf\,\, Since $f$ is a \2 mapping on $D$, for any $x \in D$, there exists  a unique $y \in D$ with $y\neq x$ satisfying $G(x,y):=f(x)-f(y)=0$, that is, the equation $G(x,y)=0$ on $\mathcal{D}$ defines a mapping $\mathcal{I}_f: x \rightarrow y$ from $D$ to $D$. Since $D$ is a finite set, by Lagrange's interpolation formula the expression of this mapping can be represented as follows, $\mathcal{I}_f(x)=\sum_{a \in D}\mathcal{I}_f(a)(1+(x+a)^{q-1})$.
Moreover, for any $a \in D$, there exists $b \in D$ with $b\neq a$ such that $G(b,a)=0$ and $G(a,b)=0$, simultaneously. So $\cI_f(a)=b$ and $\cI_f(b)=a$. That is to say, $y=\cI_f(x)$ is an involution on $D$ without any fixed point.
\EOP

\begin{rem}
We call $\cI_f(x)$ the involution derived from a \2 mapping $f$ on $D$. The following are two examples.

\noindent {\rm (1)} Let $L(x)$ be a linearized \2 mapping on $\bF_{2^n}$ and $\alpha \in \bF_{2^n}^*$ be a root of $L(x)$. For any $x \in \bF_{2^n}$, the equation $G(x,y)=L(x)+L(y)=L(x+y)=0$ has two solutions $y=x$
and $y=x+\alpha$, and so the involution derived from $L(x)$ is $\cI_L(x)=x+\alpha$.

\noindent {\rm (2)} For any APN function $F(x)$ on $\bF_{2^n}$, the derivative $D_aF(x)=F(x)+F(x+a)$ is a \2 mapping on $\bF_{2^n}$ for any $a \in \bF_{2^n}^*$. Analysis similar to that in (1) shows that the involution derived from $D_aF(x)$ is $\cI_{D_aF}(x)=x+a$.
\end{rem}


Let $D$ and $S$ be two finite sets with the same even cardinality, and let $f$ be a \2 mapping on $D$ and $p$ be a bijection from $S$ to $D$.
It is easy to see that $y=\cI_f(x)$ is the unique solution of the equation $f(x)-f(y)=0$ in  $\{(x,y) \in D^2| \,\, x \neq y\}$ if and only if $y=p^{-1}\circ \cI_f \circ p(x)$ is the unique solution of the equation $f \circ p(x)-f \circ p(y)=0$ in $\{(x,y) \in S^2| \,\, x \neq y\}$. So, we have the following proposition.

\begin{prop}\label{prop:pf}
Let $D$ and $S$ be two subsets of $\bF_q$ with the same even cardinality. Let $f$ be a \2 mapping on $D$ and $p$ be a bijection from $S$ to $D$. Then the involution derived from $f$ on $D$ is $\cI_f$ if and only if the involution derived from $f \circ p$ on $S$ is $\cI_{f\circ p}=p^{-1}\circ \cI_f \circ p$.
\end{prop}


Let $f$ be a \2 mapping, and $p$ be a injection over ${\rm Im} (f)$. We have the following proposition about the \2 mappings $p \circ f$ and $f$.

\begin{prop}\label{prop:outerbijection}
Let $A,B,C$ be three finite sets with $\sharp A=2\sharp B=2\sharp C$, $f$ be a \2 mappings from $A$ to $B$, and $\bar f$ be a \2 mapping from $A$ to $C$. Then the involutions derived from $f$ and $\bar f$ respectively both are $\cI$ if and only if there exists a bijection $p$ from $B$ to $C$ such that $\bar f=p\circ f$.
\end{prop}
{\it Proof.}  The sufficiency holds since the equations $f(x)-f(y)=0$ and $p \circ f(x)-p \circ f(y)=0$ have the same roots in $\{(x,y) \in A^2 | x \neq y\}$.

Next, we prove the necessity. Assume that the involutions derived from $f$ and $\bar f$ respectively both are $\cI$. Since $f$ is a \2 mapping over $A$, we have $A=\cup _{b \in B}A_b$, where $A_b=\{x \in A | f(x)=b\}$ and
$\sharp A_b=2$. Pick one element from each $A_b$ for $b\in B$ to form a set $\bar A$. Let $f_{\bar A}$ and $\bar{f}_{\bar A}$ denote the mappings of $f$ and $\bar f$ restricted on $\bar A$, respectively. By the definition of
$\bar A$, one can verify that $f_{\bar A}$ and $\bar{f}_{\bar A}$ are bijective from $\bar A$ to $B$ and $C$, respectively, and $\cI(\bar A)=A \setminus \bar A$.
Similarly, $f$ and $\bar f$ are also bijective from $A \setminus \bar A$ to $B$ and $C$, respectively. Hence,
$f$ and $\bar f$ can be rewritten as
\begin{equation*}
  f(x)=\left\{ \begin{aligned}
    &f_{\bar A}(x),&& x \in \bar A; \\
    &f_{\bar A}(\cI(x)),&& x \in A \setminus \bar A
       \end{aligned}\right. {\,\, \text and \,\,}
  \bar f(x)=\left\{ \begin{aligned}
    &\bar f_{\bar A}(x),&& x \in \bar A; \\
    &\bar f_{\bar A}(\cI(x)),&& x \in A \setminus \bar A.
       \end{aligned}\right.
\end{equation*}
Thus, there exists a bijection $p=\bar f_{\bar A}\circ f_{\bar A}^{-1}$ from $B$ to $C$ such that $\bar f=p\circ f$.
\EOP

For a given involution $\cI$ without fixed points over $\bF_{2^n}$, we have $\bF_{2^n}=\bigcup_{a\in \bF_{2^n}}S_a$, where $S_a=\{a, \cI(a)\}$. Pick one element from each $S_a$ for $a\in \bF_{2^n}$ to form a new set $S$.
It is easy to see that
 \begin{equation*}
  f(x)=\left\{ \begin{aligned}
    &x,&& x \in  S; \\
    &\cI(x),&& x \in \bF_{2^n} \setminus  S
 \end{aligned}\right.
 \end{equation*}
is a \2 mapping over $\bF_{2^n}$ and the involution derived from it is $\cI$. By Proposition~\ref{prop:outerbijection}, all \2 mappings which derive the involution $\cI$ can be obtained by composing $f$ with a bijection on Im$(f)$. Note that there are $\binom{2^n}{2^{n-1}}$ subsets in $\bF_{2^n}$ with cardinality $2^{n-1}$. For each such subset $T$, there are $2^{n-1}!$ bijections from $S$ to $T$. So, we have the following result.

\begin{cor}
Let $n \in \N$, for a given involution $\cI$ without fixed points on $\bF_{2^n}$, there are $\frac{2^{n}!}{2^{n-1}!}$ distinct \2 mappings from which $\cI$ can be derived.
\end{cor}

\section{AGW-like criterion for \2 mappings}\label{sec:AGW-like criterion}

The AGW criterion~\cite{AGW2011} transforms the problem of proving the permutation property of a polynomial on a finite set $A$ into showing that another polynomial is a permutation on a subset of $A$.
This technique facilitates the proof and computer experiments of finding new permutation polynomials. Recently, this method has been generalized to construct \2 mappings over $\bF_{2^n}$ by Mesnager and Qu as follows.
\begin{lem}\cite{MesnagerQu2019}\label{lem:MesnagerQu2019}
Let $A$, $S$ and $\bar S$ be finite sets with $\sharp S=\sharp \bar S$. Let $f,\bar f,\lambda, \bar \lambda$ be four mappings defined as the following diagram such that $\bar \lambda \circ f=\bar f \circ \lambda$. If $\bar f$ is a bijection from $S$ to $\bar S$, $f|_{\lambda^{-1}(s)}$ is \2 for any $s \in S$, and there is at most one $s \in S$ such that $\sharp \lambda^{-1}(s)$ is odd, then $f$ is a \2 mapping over~$A$.
\[ \begin{CD}
    A                     @>f>>    A \\
    @VV \lambda V                 @VV \bar \lambda V \\
    S                     @>\bar f >>   \bar S
    \end{CD}\]
\end{lem}
In Remark~7 of~\cite{MesnagerQu2019}, authors also wanted to derive \2 mappings $f$ on $A$ by the assumption that $\bar f$ was a \2 mapping on $S$ and $f|_{\lambda^{-1}(s)}$ was injective. But they found that it seems not easy to add a condition such that $f$ was a \2 mapping on $A$. Here, we find a restriction on $f$ and then propose another AGW-like criterion for constructions of new \2 mappings.

\begin{prop}\label{prop:AGW-like}
Let $A,\bar A,S,\bar S$ be four finite sets and $f,\bar f,\lambda, \bar \lambda$ be four surjective mappings defined as the following diagram such that $\bar \lambda \circ f=\bar f \circ \lambda$. Then $f$ is a \2 mapping on $A$ if the following two conditions hold:
\begin{enumerate}
\item[{\rm (1)}] $\bar f$ is a \2 mapping from $S$ to $\bar S$;
\item[{\rm (2)}] $\sharp S$ is even, and $f$ is bijective from $\lambda^{-1}(s)=\{ x \in A | \lambda(x)=s\}$ to $\bar \lambda^{-1}(\bar f(s))=\{ x \in \bar A | \bar \lambda(x)=\bar f(s)\}$ for any $s \in S$.
\end{enumerate}
\[ \begin{CD}
    A                     @>f>>    \bar A \\
    @VV \lambda V                 @VV \bar \lambda V \\
    S                     @>\bar f >>   \bar S
    \end{CD}\]
\end{prop}
{\it Proof.}  For any $b \in \bar A$, let $\bar s =\bar \lambda(b)$. Since $f$ is surjective, there exist some $a \in A$ such that $f(a)=b$. From above diagram we have $\bar \lambda \circ f(a)=\bar f \circ \lambda(a)=\bar s$. Since $\bar f$ is a \2 mapping from $S$ to $\bar S$, there exist exactly two elements $s_1,s_2$ in $S$ such that $\bar f(s_1)=\bar f(s_2)=\bar s$. Hence, we must have $a\in \lambda^{-1}(s_1)$ or $\lambda^{-1}(s_2)$. Note that $f$ is bijective from $\lambda^{-1}(s_i)$ to $\bar \lambda^{-1}(\bar s)$, where $i=1,2$. Thus, $b$ has exactly two preimages $a_1 \in \lambda^{-1}(s_1)$ and $a_2 \in \lambda^{-1}(s_2)$.
\EOP

By applying Proposition~\ref{prop:AGW-like}, we can propose some general constructions of \2 mappings as follows.

\begin{thm}\label{thm:relationExam1}
Let $q=2^k$ and $k,n \in \N$. Let $g, \phi, \psi, \bar \psi, h \in \bF_{q^n}[x]$ be such that $\phi$ and $\psi$ are $2$-linearized, $\bar \psi$ is $q$-linearized, $\phi\circ \psi=\bar \psi \circ \phi$, and $h(\psi(\bF_{q^n})) \subset \bF_{q}^*$. Set $f=h(\psi(x))\phi(x)+g(\psi(x))$. Then $f$ is a \2 mapping over $\bF_{q^n}$, if the following two conditions hold:
\begin{enumerate}
\item[{\rm (1)}] $\bar f=h(x)\phi(x)+\bar \psi \circ g(x)$ is a \2 mapping over $\psi(\bF_{q^n})$;
\item[{\rm (2)}] $ker (\phi)\cap ker ( \psi)=\{0\}$ and ${\rm dim \,\,} ker(\psi)={\rm dim \,\,} ker(\bar \psi)$.
\end{enumerate}
\end{thm}
{\it Proof.}
For $x \in \bF_{q^n}$, we have
\begin{align*}
  \bar \psi \circ f(x) & =\bar \psi(h(\psi(x))\phi(x))+\bar \psi(g(\psi(x))) \\
                       & =h(\psi(x))\phi(\psi(x))+\bar \psi(g(\psi(x))) \\
                       & =\bar f \circ \psi(x).
\end{align*}
The second equality holds since $h(\psi(\bF_{q^n})) \subseteq \bF_{q}^*$, $\bar \psi$ is $q$-linearized and $\phi\circ \psi=\bar \psi \circ \phi$. So, the following diagram is commutative.
\[ \begin{CD}
    \bF_{q^n}                     @>f>>    \bF_{q^n} \\
    @VV \psi V                 @VV \bar \psi V \\
    \psi(\bF_{q^n})               @>\bar f >>   \bar \psi (\bF_{q^n})
    \end{CD}\]
This shows that $f(\psi^{-1}(s)) \subseteq \bar \psi^{-1}(\bar f(s))$ for any $s \in \psi(\bF_{q^n})$.

By Proposition~\ref{prop:AGW-like}, in the following we only need to show that $f$ is bijective from $\psi^{-1}(s)$ to $\bar \psi^{-1}(\bar f(s))$ for any $s \in \psi(\bF_{q^n})$.
For some $a\neq b \in \psi^{-1}(s)$, assume that $f(a)=f(b)$, then $h(s)\phi(a-b)=0$. Note that $h(\psi(\bF_{q^n})) \subset \bF_{q}^*$. We have $a-b \in ker (\phi)\cap ker (\bar \psi)$, and so $a=b$.
This is a contradiction. Since ${\rm dim \,\,} ker(\psi)={\rm dim \,\,} ker(\bar \psi)$, $f$ is surjective from $\psi^{-1}(s)$ to $\bar \psi^{-1}(\bar f(s))$. This completes the proof.
\EOP

Let $\psi(x)=\bar \psi(x)={\rm Tr}_m^n(x)$ and $\phi(x)=x^2+x$. An explicit construction of \2 mapping is given as follows.

\begin{cor}\label{cor:involution1}
Let $m, n \in \N$ with $n/m$ odd. Let $g(x)\in \bF_{2^n}[x]$ and $h(x)=x^2+x+a \in \bF_{2^m}[x]$ with ${\rm Tr}_m(a) \neq 0$. Then $f(x)=(x^2+x)h({\rm Tr}^n_m(x))+g({\rm Tr}_m^n(x))^{2^m}+g({\rm Tr}_m^n(x))$ is a \2 mapping over $\bF_{2^n}$.
\end{cor}
{\it Proof.}  Let $\psi(x)=\bar \psi(x)={\rm Tr}_m^n(x)$, $\phi(x)=x^2+x$ and $\bar{f}(x) = h(x)\phi(x)$.
Since ${\rm Tr}_m(a) \neq 0$, $h(x)$ has no root in $\bF_{2^m}$ by Lemma~\ref{lem:square}. Since $n/m$ is odd and ${\rm Tr}^n_m(1)\neq 0$, we see that $ker (x^2+x) \cap ker ({\rm Tr}_m^n(x))=\{0\}$. Since ${\rm Tr}_m(a)\neq 0$ , by Lemma~\ref{lem:quartic}, $\bar f$ is \2 over $\bF_{2^m}$. By Theorem~\ref{thm:relationExam1}, $f(x)=(x^2+x)h({\rm Tr}^n_m(x))+g({\rm Tr}_m^n(x))^{2^m}+g({\rm Tr}_m^n(x))$ is a \2 mapping over $\bF_{2^n}$.
\EOP


\begin{thm}\label{thm:relationExam2}
Let $k,n \in \N$ and $q=2^k$. Let $L_1(x), L_2(x), L_3(x) \in \bF_{q}[x]$ be $q$-linearized, and $g(x) \in \bF_{q^n}[x]$ be such that $g(L_3(\bF_{q^n})) \subseteq \bF_{q}$. Then $f=L_1(x)+L_2(x)g(L_3(x))$ is a \2 mapping over $\bF_{q^n}$ if the following two conditions hold:
\begin{enumerate}
\item[{\rm (1)}] $\bar f(x)=L_1(x)+L_3(x)g(x)$ is a \2 mapping over $\psi(\bF_{q^n})$;
\item[{\rm (2)}] $ker (F_y)\cap ker ( L_3)=\{0\}$ for any $y \in L_3(\bF_{q^n})$, where $F_y(x)=L_1(x)+L_2(x)g(y)$.
\end{enumerate}
\end{thm}
{\it Proof.}
It is easy to verify that $L_3 \circ f=\bar f \circ L_3$, i.e., the following diagram is commutative,
\[ \begin{CD}
    \bF_{q^n}                      @>f>>    \bF_{q^n} \\
    @VV L_3 V                 @VV L_3  V \\
    L_3({\bF_{q^n}})                     @>\bar f >>   L_3(\bF_{q^n})
    \end{CD}\]
Then we see that $f(L_3^{-1}(y)) \subseteq L_3^{-1}(\bar f(y))$ for any $y \in L_3^{-1}(y)$.

 We claim that $\bar f$ is bijective from $L_3^{-1}(y)$ to $L_3^{-1}(\bar f(y))$ for any $y \in L_3(\bF_{q^n})$, if $ker (F_y)\cap ker ( L_3)=\{0\}$ for all $y \in L_3(\bF_{q^n})$. Otherwise, for some $a, b \in L_3^{-1}(y)$ with $a \neq b$, we have $f(a)=f(b)$, i.e., $F_y(a)=F_y(b)$. So $a-b \in ker (F_y)\cap ker ( L_3)$, that is, $a=b$, which is a contradiction. The proof is completed by Proposition~\ref{prop:AGW-like}.
\EOP

Let $L(x)=x^2$, $L_2(x)=x$ and $L_3(x)={\rm Tr}^{kn}_k(x)$. An more explicit construction of \2 mappings is provided as follows.
\begin{prop}
Let $q=2^k$, $k \in \N$ and $n$ be odd. Let $g(x)=x^3+bx+a \in \bF_{q}[x]$ with ${\rm Tr}_{k}(\frac{(b+1)^3}{a^2}+1)\neq 0$. Then $f=x^2+xg({\rm Tr}^{kn}_k(x))$ is a \2 mapping over $\bF_{q^n}$.
\end{prop}

{\it Proof.}  It is easy to verify that $\bar f(x)=x^2+xg(x)=x^4+(b+1)x^2+ax$ is \2 over $\bF_{2^k}$ by Lemma~\ref{lem:quartic}. For any $y \in \bF_{2^k}$, if there exists $z \in \bF_{q^n}^*$ such that $F_y(z)=z^2+zg(y)=0$ and
${\rm Tr}^{kn}_k(z)=0$, then we have $z=g(y)$, and ${\rm Tr}^{kn}_k(g(y))=ng(y)=0$, which is contradictory to that $n$ is odd, that is to say, $ker (F_y)\cap ker ( {\rm Tr}^{kn}_k(x))=\{0\}$. This completes the proof.
\EOP


By a similar analysis in Theorems~\ref{thm:relationExam1} and~\ref{thm:relationExam2}, we have the following theorem.

\begin{thm}\label{thm:relationExam3}
Let $n, k, \ell$ be positive integers with $k<n$, $\ell =\gcd(n,k)$ and $\delta\in \bF_{2^n}$. Let $S=\{ x^{2^k}+x+\delta \mid x \in \bF_{2^n} \}$, $g(x) \in \bF_{2^{n}}[x]$, $L(x)\in \bF_{2^{\ell}}[x]$ be $q$-linearized
and $h(x)=g(x)^{2^k}+g(x)+L(x)$. Then $f(x)=g(x^{2^k}+x+\delta)+L(x)$ is a \2 mapping on $\bF_{2^n}$ if the following two conditions hold:
\begin{enumerate}
\item[{\rm(1)}]   $h$ is a \2 mapping on $S$;
\item[{\rm (2)}]  $ker(L)$ $\cap\,\, ker$$(x^{2^k}+x)=\{0\}$.
\end{enumerate}
\end{thm}



\section{Two-to-one mappings of the form $(x^{2^k}+x+\delta)^s+cx$ and the involutions derived from them }\label{sec:\2 mappings}

In this section, we study the \2 property of the polynomials introduced in Theorem~\ref{thm:relationExam3}  with a more specific shape of (\ref{eq:main}),
and investigate their derived involutions over $\bF_{2^{2m}}$. In fact, the permutation property of the polynomials of the form (\ref{eq:main}) was firstly
studied by Helleseth and Zinoviev~\cite{HZ2003} to derive new Kloosterman sums identities over $\bF_{2^{n}}$, then the permutation of this type of polynomials
is specially studied by Ding and Yuan~\cite{YD2007}. Subsequently, many permutation polynomials of the form (1) or its generalization
have been obtained in the past two decades. The reader is referred to~\cite{YD2011,YD2014,ZhengYuanYu2019,Zengetal2010,TuZengJiang2015} and the references therein for information.
The following is a necessary and sufficient condition for the polynomials of the form (\ref{eq:main}) being \2 mappings.


\begin{thm}\label{thm:relation}
Let $n, k, \ell$ be positive integers with $k<n$, $\ell =\gcd(n,k)$ and $\delta\in \bF_{2^n}$. Let $S=\{ x^{2^k}+x+\delta \mid x \in \bF_{2^n} \}$ and $h(x)=x^{2^ks}+x^s+cx\in \bF_{2^{\ell}}[x]$.
Then $f(x)=(x^{2^k}+x+\delta)^s+cx$ is a \2 mapping on $\bF_{2^n}$ if and only if $h(x)$ is a \2 mapping on $S$.
\end{thm}
{\it Proof.}   The sufficiency is derived directly from Theorem~\ref{thm:relationExam3}. Next, we show the necessity.
Let $\lambda(x)=x^{2^k}+x+\delta$ and $\bar \lambda(x)=x^{2^k}+x+\delta$.  It is easy to show that the following diagram is commutative.
\[ \begin{CD}
    \bF_{2^n}                      @>f>>    \bF_{2^n} \\
    @VV \lambda V                 @VV \bar \lambda V \\
    S                     @>h >>   h(S)
    \end{CD}\]
Let $\alpha \in h(S)$, $h^{-1}(\alpha)$ and $\bar{\lambda}^{-1}(\alpha)$ denote the set of preimages of $\alpha$ under the mappings $h$ and $\bar{\lambda}$, respectively.
Let $\beta\in h^{-1}(\alpha)$ and $\lambda^{-1}(\beta)$ denote the set of preimages of $\beta$ under the mapping $\lambda$. The mapping $f$ is linear when it is restricted to $\lambda^{-1}(\beta)$, i.e.,
$f|_{\lambda^{-1}(\beta)}(x)=\beta^s+cx$. So, $f$ is a bijection from $\lambda^{-1}(\beta)$ to $\bar{\lambda}^{-1}(\alpha)$ since above diagram is commutative. Since $f$ is a 2-to-1 mapping, for any
$\alpha\in h(S)$ it has exactly two preimages under the mapping~$h$.
\EOP

\begin{prop}\label{pro:relation}
Follow the notation introduced in Theorem~\ref{thm:relation}. Let $g(x)=x^s$. If the involution derived from $h$ on $S$ is $\cI_h$, then the involution derived from $f$ on $\bF_{2^n}$ is
 $\cI_f(x)=c^{-1}(g\circ\cI_h \circ \lambda(x)+g\circ \lambda(x))+x$.
\end{prop}
{\it Proof.}
Since $f$ is a \2 mapping on $\bF_{2^n}$, for $x \in \bF_{2^n}$, there exists a unique $y \in \bF_{2^n}$ with $y\neq x$ such that
\begin{equation}\label{eq:roots}
f(x)+f(y)=g(x^{2^k}+x+\delta)+cx+g(y^{2^k}+y+\delta)+cy=0.
\end{equation}
Let $\lambda(x)=x^{2^k}+x+\delta$. Taking the both sides of ${\rm Eq.}(\ref{eq:roots})$ to the power of $2^m$, then adding it to ${\rm Eq.}(\ref{eq:roots})$, we have
\begin{equation*}
 h(\lambda(x))+h(\lambda(y))= 0.
\end{equation*}
Since the involution derived from $h$ on $S$ is $\cI_h$, we must have $\lambda(y)=\cI_h \circ \lambda(x) $ or $\lambda(x)=\lambda(y)$. If $\lambda(x)=\lambda(y)$, together with Eq.(\ref{eq:roots}), we then have $y=x$, which is a contradiction. Substituting $\lambda(y)=\cI_h \circ \lambda(x)$ into Eq.(\ref{eq:roots}), we obtain $y=c^{-1}(g\circ\cI_h \circ \lambda(x)+g\circ \lambda(x))+x$.
\EOP

\subsection{Several classes of \2 mappings and corresponding involutions over $\bF_{2^{2m}}$}

In this subsection, by application of Theorem~\ref{thm:relation} and Proposition~\ref{pro:relation} we construct several classes of \2 mappings of the form $f(x)=(x^{2^k}+x+\delta)^s+cx$  and
obtain their corresponding involutions over $\bF_{2^{2m}}$.

\begin{thm}\label{thm:lowdegree1}
Let $m\in \N$ be even and $\delta \in \bF_{2^{2m}}$ with ${\rm Tr}_{2m}(\delta)=1$. Then $f(x)=(x^2+x+\delta)^{2^{2m-2}+2^{m-2}}+x$ is a \2 mapping on $\bF_{2^{2m}}$.
\end{thm}
{\it Proof.}  It is easy to see that $p(x)=x^4$ is a permutation on $S=\{ x^2+x+\delta \mid x \in \bF_{2^{2m}}\}$. To show that $f(x)$ is 2-to-1 on $\bF_{2^{2m}}$, by Theorem~\ref{thm:relation}, it is sufficient to show that $\bar{h}(x)=h(x^4)=x^{2^{m+1}+2}+x^{2^{m}+1}+x^4$ is a \2 mapping on $S$, which can be rewritten as $\{x \in \bF_{2^{2m}}\mid {\rm Tr}_{2m}(x)=1\}$. To this end, we only need to show that for any $x\in S$, the equation
\begin{equation}\label{eq:thm2:1}
  H(y):=\bar{h}(y+x)+\bar{h}(x)=0
\end{equation}
has exactly two distinct solutions in $S_0=\{ z \in \bF_{2^{2m}} \mid {\rm Tr}_{2m}(z)=0\}$. Let $Y=y^{2^{m}}$ and $X=x^{2^{m}}$. Then Eq.(\ref{eq:thm2:1}) is reduced to
\begin{equation}\label{eq:thm2:2}
H(y)={x}^{2}{Y}^{2}+{X}^{2}{y}^{2}+{Y}^{4}+{y}^{2}{Y}^{2}+xY+Xy+Yy=0.
\end{equation}
Raising both sides of ${\rm Eq.}(\ref{eq:thm2:2})$ to the power of $2^m$ and combining $y=Y^{2^m}$ and $x=X^{2^m}$, we have
\begin{equation}\label{eq:thm2:3}
 {X}^{2}{y}^{2}+ {x}^{2}{Y}^{2}+{y}^{4}+{Y}^{2}{y}^{2}+Xy+xY+yY=0.
\end{equation}
Adding Eq.(\ref{eq:thm2:2}) to Eq.(\ref{eq:thm2:3}), we obtain
\[ (y+ Y)^4 = 0.\]
So, $Y=y$, i.e., $y \in \bF_{2^m}$. Substituting it into Eq.(\ref{eq:thm2:2}), we get
\begin{equation}\label{eq:thm2:4}
  (x^2+X^2+1)y^2+(x+X)y=0.
\end{equation}
We claim that $x+X \neq 0$ and $x+X+1 \neq 0$. Otherwise, ${\rm Tr}_{2m}(x)={\rm Tr}_{m}(x+X)={\rm Tr}_{m}(x+X+1)=0$ since $m$ is even. This contradicts to that ${\rm Tr}_{2m}(x)=1$ since $x\in S$.
Thus, Eq.(\ref{eq:thm2:4}) has exactly two solutions. Moreover, $y=\frac{x+X}{x^2+X^2+1}\in \bF_{2^m}^* \subseteq S_0$.  This completes the proof.
\EOP

\begin{cor}\label{cor:lowdegree1}
Let $m\in \N$ be even and $\delta \in \bF_{2^{2m}}$ with ${\rm Tr}_{2m}(\delta)=1$. Then $\cI(x)=x+1+\frac{1}{x^{2^{m+1}}+x^{2^m}+x^2+x+\delta+\delta^{2^m}+1}$ is an involution without any fixed point on $\bF_{2^{2m}}$.
\end{cor}
{\it Proof.}  From the proof of Theorem~\ref{thm:lowdegree1}, we see that the involution derived from $\bar{h}$ on $S$ is $\cI_{\bar h}(x)=x+\frac{x+x^{2^m}}{x^{2^{m+1}}+x^2+1}$. Since $\bar{h}= h\circ p$ and $p(x)=x^4$ is linear over $\bF_{2^{2m}}$, by Proposition~\ref{prop:pf}, the involution derived from $h$ on $S$ is
$$ \cI_{h}(x)=\cI_{\bar{h}\circ p^{-1}}(x) =  p\circ \cI_{\bar h}\circ p^{-1}(x)=\cI_{\bar h}\circ p \circ p^{-1}(x)=  \cI_{\bar h}(x).$$
Let $g(x)=x^{2^{2m-2}+2^{m-2}}$, $\lambda(x)=x^2+x+\delta$ and $c=1$. By Proposition~\ref{pro:relation}, the involution derived from $f$ in Theorem~\ref{thm:lowdegree1} is
\begin{equation*}\begin{split}
  \cI_f(x)&= g\circ \cI_{h}\circ \lambda(x)+g\circ \lambda(x)+x  \\
          &=x+1+\frac{1}{x^{2^{m+1}}+x^{2^m}+x^2+x+\delta+\delta^{2^m}+1}.
\end{split}
\end{equation*}
This completes proof.
 \EOP

\begin{thm}\label{thm:lowdegree2}
Let $m \in \N$ be even and $\delta \in \bF_{2^{2m}}$ with ${\rm Tr}_{2m}(\delta)=1$. Then $f(x)=(x^2+x+\delta)^{2^m+1}+x$ is a \2 mapping on $\bF_{2^{2m}}$.
\end{thm}
{\it Proof.}  To show that $f(x)$ is a \2 mapping on $\bF_{2^{2m}}$, by Theorem~\ref{thm:relation}, it is sufficient to prove $h(x)=x^{2^{m+1}+2}+x^{2^m+1}+x$ is a \2 mapping on $S=\{x \in \bF_{2^{2m}}\mid {\rm Tr}_{2m}(x)=1\}$. To this end, we only need to show that for any $x \in S$, the equation
\begin{equation}\label{eq:thm1:1}
\begin{split}
H(y):= &h(y+x)+h(x)=0\\
\end{split}
\end{equation}
has exactly two distinct solutions in $S_0=\{ x \in \bF_{2^{2m}} \mid {\rm Tr}_{2m}(x)=0\}$. Let $Y=y^{2^{m}}$ and $X=x^{2^{m}}$. Then Eq.(\ref{eq:thm1:1}) becomes
\begin{equation}\label{eq:thm1:2}
  y^2Y^2+x^2Y^2+X^2y^2+xY+Xy+yY+y=0.
\end{equation}
Raising both sides of Eq.(\ref{eq:thm1:2}) to the power of $2^m$ and combining $y=Y^{2^m}$ and $x=X^{2^m}$, we have
\begin{equation}\label{eq:thm1:3}
  y^2Y^2+x^2Y^2+X^2y^2+xY+Xy+yY+Y=0.
\end{equation}
Then adding Eq.(\ref{eq:thm1:2}) to Eq.(\ref{eq:thm1:3}) we get $Y=y$, i.e., $y \in \bF_{2^m}$.  Substituting it into Eq.(\ref{eq:thm1:2}), we obtain
\begin{equation}\label{eq:thm1:4}
  y^4+(x^2+X^2+1)y^2+(x+X+1)y=0.
\end{equation}
It is clear that $y=0$ is a solution of Eq.(\ref{eq:thm1:4}). Next, we show that
\begin{equation}\label{eq:thm1:5}
  y^3+(x^2+X^2+1)y+(x+X+1)=0
\end{equation}
has a unique root in $\bF_{2^m}^*$, which is a subset of $S_0$.

We claim that $x+X+1\neq 0$. Otherwise, ${\rm Tr}_{2m}(x)={\rm Tr}_{m}(x+X+1)=0$ since $m$ is even, which contradicts to that ${\rm Tr}_{2m}(x)=1$ since $x\in S$.
According to Lemma~\ref{lem:cubic}, Eq.(\ref{eq:thm1:4}) has exactly one root in $\bF_{2^{m}}^*$ since ${\rm Tr}_{m}((x^2+X^2+1)^2+1)={\rm Tr}_{m}(x+X)={\rm Tr}_{2m}(x)=1$.  \EOP

By a similar argument as above, we have the following theorem.
\begin{thm}\label{thm:lowdegree3}
Let $m \in \N$ be even, and $\delta \in \bF_{2^{2m}}$ with ${\rm Tr}_{2m}(\delta)=1$. Then $f(x)=(x^2+x+\delta)^{2^{2m-1}+2^{m-1}}+x$ is a \2 mapping on $\bF_{2^{2m}}$.
\end{thm}

{\rem Since the explicit expression of the involution derived from $h$ on $S$ is not clear, we can't obtain the explicit expression of the involutions derived from the \2 mappings in Theorems~\ref{thm:lowdegree2} and \ref{thm:lowdegree3}.}

\begin{thm}\label{thm:squareexponent}
Let $m, i \in \N$ with ${\rm gcd}(m,i)=1$ and $\delta \in \bF_{2^{2m}}$, $c \in \bF_{2^m}^*$ with ${\rm Tr}_{m}^{2m}(\delta^2+c^{2^{m-i}}\delta)\neq 0$. Then $f(x)=(x^{2^m}+x+\delta)^{2^i+1}+cx$ is a \2 mapping on $\bF_{2^{2m}}$.
\end{thm}
{\it Proof.}   To show that $f(x)$ is a \2 mapping on $\bF_{2^{2m}}$, by Theorem~\ref{thm:relation}, it is sufficient to show that $h(x)=x^{(2^i+1)\cdot 2^m}+x^{2^i+1}+cx$ is a \2 mapping on $S=\{x  \in \bF_{2^{2m}}\mid {\rm Tr}^{2m}_{m}(x) = {\rm Tr}^{2m}_{m}(\delta)\}$. To this end, we only need to show that for any $x\in S$, the equation
\begin{equation}\label{eq:thm3:1}
\begin{split}
H(y):= &h(y+x)+h(x)=0
\end{split}
\end{equation}
has exactly two distinct solutions in $S_0=\{ z \in \bF_{2^{2m}}\mid {\rm Tr}^{2m}_{m}(z)=0\}$, which is exactly $\bF_{2^{m}}$.
Let $X=x^{2^m}$. Since $y^{2^m}=y$ for $y\in \bF_{2^m}$, Eq.(\ref{eq:thm3:1}) is reduced to
\begin{equation*}\label{eq:thm3:3}
 H(y)=(x+X)y^{2^i}+(x^{2^i}+X^{2^i}+c)y=0.
\end{equation*}
Next, we show that
\begin{equation}\label{eq:thm3:4}
  (x+X)y^{2^i-1}+x^{2^i}+X^{2^i}+c=0
\end{equation}
has a unique root in $\bF_{2^{m}}^*$.

Let $\gamma={\rm Tr}^{2m}_{m}(\delta)$. From $x\in S$ we see that~$x+X=\gamma$. Since ${\rm Tr}_{m}^{2m}(\delta^2+c^{2^{m-i}}\delta)\neq 0$, i.e.,
$\gamma^2+c^{2^{m-i}}\gamma \neq 0$, we have $x+X\neq 0$ and $x^{2^i}+X^{2^i}+c \neq 0$. Thus, Eq.(\ref{eq:thm3:4}) has a unique root $y=(\gamma^{2^i-1}+\frac{c}{\gamma})^{\frac{1}{2^i-1}}$ in $\bF_{2^{m}}^*$, where $\frac{1}{2^i-1}$ denotes the inverse of $2^i-1$ modulo $2^m-1$ since ${\rm gcd}(m,i)=1$.
\EOP

\begin{thm}\label{thm:cubicexponent}
Let $m,i \in \N$, $\delta \in \bF_{2^{2m}}$, $c \in \bF_{2^m}$ and ${\rm Tr}_{m}^{2m}(\delta) = \gamma$ with $\gamma^{2^i+2}+c\gamma \neq 0$. Then $f(x)=(x^{2^m}+x+\delta)^{2^m+2^i+1}+cx$ is a \2 mapping on $\bF_{2^{2m}}$.
\end{thm}

{\it Proof.}  To show that $f(x)$ is a \2 mapping on $\bF_{2^{2m}}$, by Theorem~\ref{thm:relation}, it is sufficient to show that $h(x)=x^{2^m+2^i+1}+x^{1+(2^i+1)\cdot 2^m}+cx$ is a \2 mapping on $S=\{x  \in \bF_{2^{2m}}\mid {\rm Tr}^{2m}_{m}(x) = {\rm Tr}^{2m}_{m}(\delta)\}$. To this end, we only need to show that for any $x \in S$, the equation
\begin{equation}\label{eq:thm4:1}
\begin{split}
H(y):= &h(y+x)+h(x)
\end{split}
\end{equation}
has exactly two distinct solutions in $S_0=\{ x \in \bF_{2^{2m}} \mid {\rm Tr}^{2m}_{m}(x)=0\}$, which is exactly $\bF_{2^{m}}$. Let $X=x^{2^m}$. Since $y^{2^m}=y$ for $y\in \bF_{2^m}$, Eq.(\ref{eq:thm4:1}) is reduced to
\begin{equation}\label{eq:thm4:3}
   ( {x}^{{2}^{i}}+{X}^{{2}^{i}} ) y^2+((x+X)^{2^i+1}+c)y=0.
\end{equation}
Next, we only need to show that
\begin{equation}\label{eq:thm4:4}
  ( {x}^{{2}^{i}}+{X}^{{2}^{i}} ) y+(x+X)^{2^i+1}+c=0
\end{equation}
has a unique root in $\bF_{2^{m}}^*$.

Since $\gamma^{2^i+2}+c\gamma\neq 0$, we have $x^{2^i}+X^{2^i}\neq 0$ and $(x+X)^{2^i+1}+c \neq 0$. Thus, Eq.(\ref{eq:thm4:4}) has a unique root $y=\gamma+\frac{c}{\gamma^{2^i}}$ in $\bF_{2^{m}}^*$.
\EOP

\begin{rem}\label{rem:lininvolution}
By Proposition~\ref{prop:pf}, if the involution derived from the \2 mapping $h(x)=x^{2^ks}+x^s+cx$ on $S$ is $\cI_h(x)=x+\xi$, where $\xi \in \bF_{2^m}^*$ is a constant. Then the involution derived from $f(x)=(x^{2^k}+x+\delta)^s+cx$ on $\bF_{2^{2m}}$ is $\cI_{f}(x)=c^{-1} [(x^{2^k}+x+\delta+\xi)^s+(x^{2^k}+x+\delta )^s ]+x$. As a consequence, the involution derived from the \2 mapping in Theorem~\ref{thm:squareexponent} is linear. The involution derived from \2 mapping in Theorem~\ref{thm:cubicexponent} is given in the following corollary.
\end{rem}

\begin{cor}\label{cor:cubicexponent}
Let $m, i \in \N$, $\delta \in \bF_{2^{2m}}$, $c \in \bF_{2^m}^*$  and $\gamma=\delta+\delta^{2^m}$ with $\gamma^{2^i+2}+c\gamma \neq 0$. Then $\cI(x)=c^{-1}[(x^{2^m}+x+\delta+\gamma+\frac{1}{\gamma^{2^i}})^{2^m+2^i+1}+(x^{2^m}+x+\delta)^{2^m+2^i+1} ]+x$ is an involution without any fixed point on $\bF_{2^{2m}}$.
\end{cor}

%

\begin{thm}\label{thm:factionalexponent}
Let $m \in \N$ be even and $\delta \in \bF_{2^{2m}}$ with ${\rm Tr}_{2m}(\delta)=1$. Then $f(x)=(x^{2}+x+\delta)^{\frac{2^{2m}+2^m+1}{3}}+x$ is a \2 mapping on $\bF_{2^{2m}}$.
\end{thm}
{\it Proof.}  To show that $f(x)$ is a \2 mapping on $\bF_{2^{2m}}$, by Theorem~\ref{thm:relation}, it is sufficient to show that $h(x)=x^{\frac{2^{2m}+2^m+1}{3}}+x^{\frac{2 (2^{2m}+2^m+1)}{3}}+x$ is a \2 mapping on $S=\{x  \in \bF_{2^{2m}}\mid {\rm Tr}_{2m}(x)=1\}$. Note that $(2-2^m)\cdot\frac{2^{2m}+2^m+1}{3} \equiv 1 \pmod{2^{2m}-1}$. So, $p(x)=x^{2-2^m}$ is a permutation on $\bF_{2^{2m}}^*$, and it is also a bijection from $\bar S=\{ x \in \bF_{2^{2m}}\mid {\rm Tr}_{2m}(x^{2-2^m})=1\}$ to $S$. Now, we only need to prove $h(x^{2-2^m})$ is a \2 mapping on $\bar S$. Next, we show the equation
\begin{equation}\label{eq:factionalexponent}
h(y^{2-2^m})-h(x^{2-2^m}) = y^2+y+y^{2-2^m}-x^2+x+x^{2-2^m}=0
\end{equation}
has a unique solution $y\in \bar{S}$ with $y\neq x$, where $x\in \bar{S}$.

Let $X=x^{2^m}$, $Y=y^{2^m}$. Substituting these into Eq.(\ref{eq:factionalexponent}) and then raising both sides of Eq.(\ref{eq:factionalexponent}) to the power of $2^m$, we have
\begin{equation}\label{eq:factionalexponent2}
\left\{ \begin{gathered}
  x^2+x+\frac{x^2}{X}=y^2+y+\frac{y^2}{Y},\\
  X^2+X+\frac{X^2}{x}=Y^2+Y+\frac{Y^2}{y}.
\end{gathered} \right.
\end{equation}
The system of Eq.(\ref{eq:factionalexponent2}) is equivalent to the following system,
\begin{equation*}
\left\{ \begin{aligned}
  &P(x,X,y,Y):= \left( {y}^{2}X+Xy+{x}^{2}+xX+{x}^{2}X \right) Y+{y}^{2}X=0,\\
  &Q(x,X,y,Y):= \left( xy+x \right) {Y}^{2}+Yxy+{X}^{2}xy+yxX+y{X}^{2}=0.
\end{aligned} \right.
\end{equation*}
Since ${\rm Tr}_{2m}(x^2+x+\frac{ x^2}{X})={\rm Tr}_{2m}(\frac{x^2}{X})={\rm Tr}_{2m}(x^{2-2^m})=1$, by Lemma~\ref{lem:square}, there is no $y\in \bF_{2^{2m}}$
satisfying ${y}^{2}X+Xy+{x}^{2}+xX+{x}^{2}X = 0$ for $x,X \in \bar S$, i.e., ${y}^{2}X+Xy+{x}^{2}+xX+{x}^{2}X \neq 0$ for any $x, y\in \bar S$. Computing the resultant of
$P$ and $Q$ with respect to $Y$ gives
\begin{equation}\label{eq:res}
Res(P,Q,Y)=xX(x+y)^2(xX+x+X)(Xy+xX+x+X)^2=0.
\end{equation}
We claim that $xX+x+X \neq 0$ for $x \in \bar S$. Otherwise, $(x  X+ x+ X)\frac{ x}{X}=0$, and then ${\rm Tr}_{2m}((x  X+ x+ X)\frac{ x}{ X})={\rm Tr}_{2m}( x^2+ x+\frac{ x^2}{X})={\rm Tr}_{2m}( x^{2-2^m})=0$. This is a contradiction. So, Eq.(\ref{eq:res}) has two solutions $y=x$ and $y=x+1+\frac{x}{X}$.
It remains to show that the second solution is in $\bar S$.
\[ \begin{split}
 {\rm Tr}_{2m}(y^{2-2^m}) &={\rm Tr}_{2m}\left(\left(\frac{xX+x+X}{X}\right)^{2-2^m}\right) \\
&={\rm Tr}_{2m}\left(\frac{x^2}{X}+\frac{x^2}{X^2}+\frac{x}{X}\right) ={\rm Tr}_{2m}(x^{2-2^m})=1.
\end{split}
\]
Thus, $y=x+1+\frac{x}{X}\in \bar{S}$ is the unique solution of Eq.(\ref{eq:factionalexponent}) with $y\neq x$.
\EOP

\begin{cor}\label{cor:factionalexponent}
Let $m \in \N$ be even and $\delta \in \bF_{2^{2m}}$ with ${\rm Tr}_{2m}(\delta)=1$. Then $\cI(x)=(x^2+x+\delta)^{\frac{2^{2m+1}-2^m-1}{3}}+x+1$ is an involution without any fixed point in $\bF_{2^{2m}}$.
\end{cor}
{\it Proof.}  Let $g(x)=x^{\frac{2^{2m}+2^m+1}{3}}$, $c=1$. From Theorem~\ref{thm:factionalexponent}, we see that the involution derived from $h\circ g^{-1}$ on $\bar S$ is $\cI_{h\circ g^{-1}}(x)=x+1+{x^{1-2^m}}$. By Proposition~\ref{prop:pf}, the involution derived from $h$ on $S$ is $\cI_h=g^{-1}\circ \cI_{h\circ g^{-1}} \circ g$. Let $\lambda(x)=x^2+x+\delta$ and $c=1$, then by Proposition~\ref{pro:relation}, the involution derived from $f$ on $\bF_{2^{2m}}$ is
\[ \begin{split}
&\cI_f(x)=g\circ\cI_h \circ \lambda(x)+g\circ \lambda(x)+x =\cI_{h\circ g^{-1}} \circ g \circ \lambda(x)+g\circ \lambda(x)+x\\
&=(x^2+x+\delta)^{\frac{2^{2m}+2^m+1}{3}(1-2^m)}+x+1=(x^2+x+\delta)^{\frac{2^{2m+1}-2^m-1}{3}}+x+1.
\end{split}
\]
\EOP


In the following, we consider the \2 mappings of the form~(\ref{eq:main}) with $s$ being a Niho exponent. The \2 mappings of this type are connected to the polynomials
of the form $h(x)=x^r\bar{h}(x^s)$ over $\bF_{2^n}$, where $s \,|\, (2^n-1)$ and $\bar{h}[x] \in \bF_{2^n}[x]$. It is known that the permutation property of the polynomials
of this form have been extensively studied in the literature. See, for example, \cite{AGW2011,DingQuWangYuanYuan2015,GuptaSharma2016,Hou2013-2014,Kyureghyan-Zieve2016,LiHelleseth12016,LiQuChen2017,LiQuWang2017,TuZengHu2014,Wang2007,ZhaHuFan2017,
Zieve2009,Zieve2013,ZhengYuanYu2019}. To this end, we first give some preliminary lemmas.

\begin{lem}\label{lem:relation}
Let $m \in \N$, $\delta \in \bF_{2^{2m}}\setminus \bF_{2^{m}}$ and $ \bar h(x)\in \bF_{2^{2m}}[x]$. Then $h(x)=x^r \bar h(x^{2^m-1})$ is a \2 mapping on $S=\{ z+\delta \mid z \in \bF_{2^{m}} \}$ if and only if $\varphi(x)=x^r \bar h(x)^{2^m-1}$ is a \2 mapping on $\mu_{2^m+1}^*$.
\end{lem}
{\it Proof.}  Let $\phi(x)=x^{2^m-1}$. It is easy to verify that $\phi \circ h=\varphi \circ \phi$. Moreover, we show that $\phi$ is a bijection from $S$ to $\mu_{2^m+1}^*$. For any $z \in \bF_{2^m}$, we have
 \[  \left( \phi(z+\delta)\right)^{2^m+1}= \left( (z+\delta)^{2^m-1}\right)^{2^m+1}=1.\]
So, $\phi(z+\delta) \in \mu_{2^m+1}^*$. If there exist $z_1, z_2 \in \bF_{2^m}$ such that $\phi(z_1+\delta)=\phi(z_2+\delta)$, i.e.,
\[ \frac{z_1+\delta^{2^m}}{z_1+\delta}=\frac{z_2+\delta^{2^m}}{z_2+\delta}. \]
This implies that $z_1=z_2$ since $\delta \in \bF_{2^{2m}}\setminus \bF_{2^{m}}$. Then the result follows from the fact that $h=\phi^{-1}\circ \varphi \circ \phi$.
\EOP


\begin{lem}\label{lem:mapping}
Let $m \in \N$ and $\theta \in \mu_{2^m+1}^*$. Then $\psi(x)=\frac{1+\theta x}{\theta+x}$ is a bijection from $\mu_{2^m+1}^* \setminus \{\theta\}$ to $\bF_{2^m} \setminus \{1\}$.
\end{lem}
{\it Proof.}  Note that $\psi(x)^{2^m}=\frac{1+(\theta x)^{-1}}{\theta^{-1}+x^{-1}}=\psi(x)$ if $x \in \mu_{2^m+1}^* \setminus \{\theta\}$, and $\psi(x)=1$ if and only if $x=1$. So, $\psi(x) \in \bF_{2^m} \setminus \{1\}$ for $x \in \mu_{2^m+1}^* \setminus \{\theta\}$. If there exist two elements $z_1, z_2 \in \mu_{2^m+1}^*$ such that $\psi(z_1)=\psi(z_2)$, i.e., $\frac{1+\theta z_1}{\theta+z_1}=\frac{1+\theta z_2}{\theta +z_2}$. Then we have  $(\theta^2+1)(z_1+z_2)=0$, and so $z_1=z_2$. This shows that $\psi$ is a bijection from $\mu_{2^m+1}^* \setminus \{\theta\}$ to $\bF_{2^m} \setminus \{1\}$ since these two sets have the same cardinality.
\EOP

\begin{thm}\label{thm:niho1}
Let $m \in \N$, $\delta \in \bF_{2^{2m}} \setminus \bF_{2^m}$ and $c \in \bF_{2^m}^*$ with ${\rm Tr}_{m}(\frac{1}{c}+1)=1$. Then $f(x)=(x^{2^m}+x+\delta)^{2^{2m-2}+2^m-2^{m-2}}+cx$ is a \2 mapping on $\bF_{2^{2m}}$.
\end{thm}
{\it Proof.}   To show that $f$ is a \2 mapping on $\bF_{2^{2m}}$, by Theorem~\ref{thm:relation}, it is sufficient to show that
\begin{align*}
h(x)&=x^{2^{2m-2}+2^m-2^{m-2}}+x^{(2^{2m-2}+2^m-2^{m-2})2^m}+cx \\
    &=x^{(2^{m-2}+1)(2^m-1)+1}+x^{(2^m-2^{m-2}+1)(2^m-1)+1}+cx
\end{align*}
is a \2 mapping on $S=\{ z +\delta \mid z \in \bF_{2^{m}}\}$. By Lemma~\ref{lem:relation}, we only need to show $\varphi(x)=x(x^{2^{m-2}+1}+x^{2^m-2^{m-2}+1}+c)^{2^m-1}$ is a \2 mapping on $\mu_{2^m+1}^*$.
First, we show that there is no $x\in \mu_{2^m+1}$ such that $x^{2^{m-2}+1}+x^{2^m-2^{m-2}+1}+c=0$. Otherwise, suppose $x\in \mu_{2^m+1}^*$ satisfies
\begin{equation}\label{eq:niho1}
 x^{2^{m-2}+1}+x^{-2^{m-2}}+c=0.
\end{equation}
Multiplying the both sides of Eq.(\ref{eq:niho1}) by $x^{-1}$, then taking them to the power of $2^m$ and combining the facts $x^{2^{2m}}= x$ and $x^{2^m}=x^{-1}$, we have
\begin{equation}\label{eq:niho2}
x^{-2^{m-2}}+x^{2^{m-2}+1}+cx=0.
\end{equation}
From Eq.(\ref{eq:niho1}) and Eq.(\ref{eq:niho2}) we have $x=1$. This is a contradiction to $x \in \mu_{2^m+1}^*$.

To show that $\varphi(x)$ is a \2 mapping on $\mu_{2^m+1}^*$, we next prove that $\varphi(x^4)$ is a \2 mapping on $\mu_{2^m+1}^*$ since ${\rm gcd}(4, 2^m+1)=1$.
Assume that there are two distinct elements $x, y \in \mu_{2^m+1}^*$ such that $\varphi(x^4)=\varphi(y^4)$, i.e.,
\begin{equation}\label{eq:niho3}
\frac{cx^4+x^3+x}{x^3+x+c}=\frac{cy^4+y^3+y}{y^3+y+c}.
\end{equation}
Since $x\neq y$, Eq.(\ref{eq:niho3}) is reduced to
\begin{equation}\label{eq:niho4}
x^3y^3+ x^2y^2+ xy +1+ x^2+y^2+ x^3y+xy^3+ c(x+y)^3=0.
\end{equation}
Eq.(\ref{eq:niho4}) is further reduced to
\begin{equation}\label{eq:niho5}
z^3 + z + c=0,
\end{equation}
where $z=\frac{1+xy}{x+y}\in \bF_{2^m}$. Since ${\rm Tr}_{m}(\frac{1}{c}+1)=1$, by Lemma~\ref{lem:cubic}, we know that Eq.(\ref{eq:niho5}) has a unique
solution $z=\alpha \in \bF_{2^{m}} \setminus \{1\}$. Thus, by Lemma~\ref{lem:mapping}, $y=\frac{\alpha x+1}{x+\alpha}$ is the unique element in $\mu_{2^m+1}^*\setminus \{x\}$ satisfying $\varphi(x^4)=\varphi(y^4)$. This completes the proof.
\EOP

\begin{cor}\label{cor:niho1}
Let $m \in \N$, $\delta \in \bF_{2^{2m}}\setminus \bF_{2^m}$ and $c \in \bF_{2^m}^*$ with ${\rm Tr}_{m}(\frac{1}{c}+1)=1$. Let $\alpha\in \bF_{2^m}$ be the root of $x^3+x+c=0$. Then
\[\begin{split} \cI_f(x)=c^{-1}[(x^{2^m}+x+\delta+&\frac{(\delta+\delta^{2^{m}})}{\alpha^4+1})^{2^{2m-2}+2^m-2^{m-2}}\\
                                                                     &+(x^{2^m}+x+\delta)^{2^{2m-2}+2^m-2^{m-2}}]+x \end{split}\]
is an involution without any fixed point in $\bF_{2^{2m}}$.
\end{cor}
{\it Proof.}  Follow the notation introduced in Theorem~\ref{thm:niho1} and let $\phi(x)=x^{2^m-1}$, $p(x)=x^4$.
By Theorem~\ref{thm:niho1}, the involution derived from $\varphi\circ p$ on $\mu_{2^m+1}^*$ is $\cI_{\varphi\circ p}(x)=\frac{\alpha x+1}{x+\alpha}$. Then, by Theorem~\ref{prop:pf}, the involution derived from $\varphi$ is $\cI_\varphi(x)=p\circ \cI_{\varphi\circ p} \circ p^{-1}(x) =\frac{\alpha^4 x+1}{x+\alpha^4}$.  For any $x\in S$, there exists $z\in \bF_{2^m}$ such that $x= z+\delta$.
Then $\phi(x)=x^{2^m-1}=\frac{z+\delta^{2^m}}{z+\delta}=1+\frac{\delta+\delta^{2^m}}{x}$ and $\phi^{-1}(x) = \frac{\delta+\delta^{2^m}}{1+x}$. By Theorem~\ref{prop:pf} again, the involution derived from $h$ on $S$ is $\cI_h(x)=\phi^{-1}\circ \cI_\varphi \circ \phi(x)=x+\frac{\delta+\delta^{2^m}}{\alpha^4+1}$ since $h(x)=\phi^{-1} \circ \varphi\circ \phi(x)$. Thus, by Remark~\ref{rem:lininvolution}, the involution derived from $f$ on $\bF_{2^{2m}}$ is $\cI_f(x)=c^{-1}[(x^{2^m}+x+\delta+\frac{(\delta+\delta^{2^{m}})}{\alpha^4+1})^{2^{2m-2}+2^m-2^{m-2}}+(x^{2^m}+x+\delta)^{2^{2m-2}+2^m-2^{m-2}}]+x$.
\EOP

\begin{thm}\label{thm:niho2}
Let $m \in \N$, $\delta \in \bF_{2^{4m}} \setminus \bF_{2^{2m}}$ and $c \in \bF_{2^{2m}} \setminus \bF_{2^m} $. Then $f(x)=(x^{2^{2m}}+x+\delta)^{(2^{2m-1}-2^{m-1}+1)(2^{2m}-1)+1}+cx$ is a \2 mapping on $\bF_{2^{4m}}$.
\end{thm}
{\it Proof.}   To show that $f$ is a \2 mapping on $\bF_{2^{4m}}$, by Theorem~\ref{thm:relation}, it is sufficient to show that
\begin{equation*}
h(x)=x^{(2^{2m-1}-2^{m-1}+1)(2^{2m}-1)+1}+x^{(2^{m-1}-2^{2m-1})(2^{2m}-1)+1}+cx
\end{equation*}
is a \2 mapping on $S= \{z+\delta \mid z \in \bF_{2^{2m}}\}$. By Lemma~\ref{lem:relation}, we only need to show $\varphi(x)=x(x^{2^{2m-1}-2^{m-1}+1}+x^{2^{m-1}-2^{2m-1}}+c)^{2^{2m}-1}$ is a \2 mapping on $\mu_{2^{2m}+1}^*$. First, we show that for any $x\in \mu_{2^{2m}+1}^*$,
$ x^{2^{2m-1}-2^{m-1}+1}+x^{2^{m-1}-2^{2m-1}}+c\neq 0.$
Otherwise, assume that $x_0 \in \mu_{2^{2m}+1}^*$ satisfies
\begin{equation}\label{eq:niho21}
x_0^{2^{2m-1}-2^{m-1}+1}+x_0^{2^{m-1}-2^{2m-1}}+c=0.
\end{equation}
Multiplying the both sides of Eq.(\ref{eq:niho21}) by $x_0^{-1}$, then taking them to the power of $2^{2m}$, we get
\begin{equation}\label{eq:niho22}
x_0^{2^{m-1}-2^{2m-1}}+x_0^{2^{2m-1}-2^{m-1}+1}+cx_0=0.
\end{equation}
From Eq.(\ref{eq:niho21}) and Eq.(\ref{eq:niho22}) we have $c(x_0+1)=0$. This contradicts with that $x_0 \in \mu_{2^{2m}+1}^*$.

Next, we show that $\varphi(x^2)$ is a \2 mapping on $\mu_{2^{2m}+1}^*$ since ${\rm gcd}(2, 2^{2m}+1)=1$. Assume that there exist $x, y \in \mu_{2^{2m}+1}^*$ with $x\neq y$ such that $\varphi(x^2)=\varphi(y^2)$, i.e.,
\begin{equation}\label{eq:niho23}
\frac{x^{2^{2m+1}-2^{m+1}+2}+cx^{2^{2m}-2^{m}+2}+1}{x^{2^{2m+1}-2^{m+1}+2}+cx^{2^{2m}-2^m}+1}= \frac{y^{2^{2m+1}-2^{m+1}+2}+cy^{2^{2m}-2^{m}+2}+1}{y^{2^{2m+1}-2^{m+1}+2}+cy^{2^{2m}-2^m}+1}.
\end{equation}
Let $X=x^{2^m}$ and $Y=y^{2^m}$. Note that $x^{2^{2m}+1} = y^{2^{2m}+1}=1$ and from Eq.(\ref{eq:niho23}) we have
\begin{equation}\label{eq:niho24}
\begin{split}
&P(x,X,y,Y):= \left( y{x}^{2}X+Xy \right) {Y}^{2} \\
&+ \left( {y}^{2}x{X}^{2}+{y}^{2}cX+c{x}^{2}X+{y}^{2}x+{X}^{2}x+x \right) Y+y{x}^{2}X+Xy=0.
\end{split}
\end{equation}
Let $\bar c=c^{2^m}$. Note that $Y^{2^m}=y^{-1}$ and $X^{2^m}=x^{-1}$. Taking the both sides of Eq.(\ref{eq:niho24}) to the power of $2^{m}$, then multiplying them by $x^2y^2$, we have
\begin{equation}\label{eq:niho25}
\begin{split}
Q(x,X,y,Y)&:= x^4y^4P(x,X,y,Y)^{2^{m}}+ \left( {y}^{2}x{X}^{2}+{y}^{2}x+{X}^{2}x+x \right) Y \\
&+\bar c{X}^{2}yx+y{x}^{2}X+Xy=0.
\end{split}
\end{equation}
By calculating the resultant of $P$ and $Q$ with respect to $Y$, we have
\begin{equation}\label{eq:resultant}
Res(P, Q, Y) = y^2X^2(y+x)^2(A(x,X)y^2+B(x,X))=0,
\end{equation}
where $A(x, X)=A_1(x,X)A_2(x, X)$, $B(x, X)=B_1(x,X)B_2(x, X)$ and
\begin{equation*}\label{eq:AB}
\begin{gathered}
A_1(x,X)={x}^{2}{X}^{2}\bar c+xXc\bar c+{x}^{2}\bar c+c{x}^{2}+c , \\
A_2(x,X)={x}^{2}{X}^{2}\bar c+c{x}^{2}{X}^{2}+xXc\bar c+{x}^{2}\bar c+c{X}^{2},\\
B_1(x,X)=x^2(xXc\bar c+\bar c{X}^{2}+c{x}^{2}+\bar c+c) ,\\
B_2(x,X)=c{x}^{2}{X}^{2}+xXc\bar c+\bar c{X}^{2}+c{X}^{2}+\bar c .
\end{gathered}
\end{equation*}
In the following we show that $A_1(x, X)\neq 0$ for any $x\in\mu_{2^{2m}+1}^*$. And $A_2(x,X)\neq0$, $B_1(x,X)\neq0$, $B_2(x,X)\neq0$, $A(x,X)+B(x,X)\neq 0$
and $A(x,X)+x^2 B(x,X)\neq 0$  for $x\in\mu_{2^{2m}+1}^*$ can be proved similarly. Assume that there exists $\bar{x} \in \mu_{2^{2m}+1}^*$ such that $A_1(\bar{x}, \bar{X})=0$, where $\bar X=\bar{x}^{2^m}$. Then
$$ \bar{x}^2\bar{X}^2A_1(\bar{x}, \bar{X})^{2^m}+A_1(\bar{x}, \bar{X})=(\bar c+c)(\bar{x}+ \bar{X})^2=0.$$
 This is a contradiction since $c \in \bF_{2^{2m}}\setminus \bF_{2^m}$ and $\bar{x} \in \mu_{2^{2m}+1}^*$. For $x \in \mu_{2^{2m}+1}^*$, it is easy to show that
 \[\left(\frac{B(x,X)}{A(x,X)}\right)^{2^{2m}}=\frac{B(\frac{1}{x},\frac{1}{X})}{A(\frac{1}{x},\frac{1}{X})} =\frac{A(x,X)}{B(x,X)}. \]
This shows that $\frac{B(x,X)}{A(x,X)} \in \mu_{2^{2m}+1}^*$. Thus, from Eq.(\ref{eq:resultant}) we know that $y=\sqrt{\frac{B(x,X)}{A(x,X)}}$ is the unique solution
of Eq.(\ref{eq:niho23}) with $y\neq x$.
\EOP

\begin{cor}\label{cor:niho2}
Let $m \in \N $, $\delta \in \bF_{2^{4m}} \setminus \bF_{2^{2m}}$, $c \in \bF_{2^{2m}} \setminus \bF_{2^m} $, and let $A,B\in \bF_{2^{4m}}[x,X]$ be defined in Eq.(\ref{eq:resultant}), and let $g(x)=x^{(2^{2m-1}-2^{m-1}+1)(2^{2m}-1)+1}$, $\phi(x)=x^{2^{2m}-1}$, $\lambda(x)=x^{2^{2m}}+x+\delta$ and $\cI_h(x)=\phi^{-1} \circ \frac{B(\sqrt{x},x^{2^{m-1}})}{A(\sqrt{x},x^{2^{m-1}})} \circ \phi(x)$. Then $\cI(x)=c^{-1}(g\circ \cI_h \circ \lambda(x)+g\circ \lambda(x))+x$ is an involution on $\bF_{2^{4m}}$.
\end{cor}

All results in this section are summarized in Table~1.

\begin{table}[h]\label{tab:1}
\caption{\rm  \2 mappings of the form ${(x^{2^k}+x+\delta)}^s+cx\in \bF_{2^{2m}}[x]$ }
\begin{center}
\resizebox{\textwidth}{!}{
\begin{tabular}{ccccccc}\hline
  No &  $k$ &  $m$ & $s$  &  Conditions     &Ref.     &  Involution \\
  \hline
  1  &  $1$ &  even & $2^m+1$   & ${\rm Tr}_{2m}(\delta)=1, c=1$   &  Thm.~\ref{thm:lowdegree2} & \\

  2  &  $1$ &  even & $2^{2m-1}+2^{m-1}$     & ${\rm Tr}_{2m}(\delta)=1, c=1$        &   Thm.~\ref{thm:lowdegree3} &  \\

  3  &  $1$ &  even & $2^{2m-2}+2^{m-2}$     & ${\rm Tr}_{2m}(\delta)=1, c=1$       &     Thm.~\ref{thm:lowdegree1} & Cor.~\ref{cor:lowdegree1}\\

  4  &  $1$ &  even & $\frac{2^{2m}+2^m+1}{3}$     & ${\rm Tr}_{2m}(\delta)=1, c=1$  &     Thm.~\ref{thm:factionalexponent} & Cor.~\ref{cor:factionalexponent}\\

  5  &  $m$ &  ${\rm gcd}(m,i)=1$  & $2^i+1$   & ${\rm Tr}^{2m}_m(\delta^2+c^{2^{m-i}}\delta) \neq 0$& Thm.~\ref{thm:squareexponent}& affine polynomial\\

  6  &  $m$  & $m\in \N$  & $2^{m}+2^i+1$ & ${\rm Tr}^{2m}_m(\delta)^{2^i+2}+c{\rm Tr}^{2m}_m(\delta) \neq 0$ & Thm.~\ref{thm:cubicexponent}& Cor.~\ref{cor:cubicexponent} \\

  7  &  $m$  & $m\in\N$  & $2^{2m-2}+2^m-2^{m-2}$ & $\delta \in \bF_{2^{2m}} \setminus \bF_{2^{m}},{\rm Tr}(\frac{1}{c}+1)\neq 0,c \in \bF_{2^m}^*$ & Thm.~\ref{thm:niho1}& Cor.~\ref{cor:niho1}\\

  8  &  $m$  &  even  & $(2^{m-1}-2^{m/2-1}+1)(2^m-1)+1$ & $\delta \in \bF_{2^{2m}} \setminus \bF_{2^{m}},c \in \bF_{2^{m}} \setminus \bF_{2^{m/2}} $ & Thm.~\ref{thm:niho2}& Cor.~\ref{cor:niho2}\\
  \hline
\end{tabular}}
\end{center}
\end{table}

\section{The involutions derived from the known \2 mappings over $\bF_{2^{2m+1}}$}\label{sec:involution for odd n}

By using Theorem~\ref{thm:main} and the multivariate technique introduced by Dobbertin~\cite{Dobbertin2002}, we obtain five classes of new involutions on $\bF_{2^{2m+1}}$
from known \2 mappings proposed in~\cite{LiMesnagerQu2019}. To this end, we first recall the known \2 mappings over $\bF_{2^{2m+1}}$ introduced in~\cite{LiMesnagerQu2019}.
\begin{lem}\label{lem:2-1}\cite{LiMesnagerQu2019}
For $m\in \N$, the polynomial $f(x)$ is a \2 mapping over $\bF_{2^{2m+1}}$ in the following cases:
\begin{enumerate}
\item[{\rm (1)}] $f(x)=x^{2^{m+1}+2}+x^{2^{m+1}}+x^2+x$;
\item[{\rm (2)}] $f(x)=x^{2^{m+1}+2}+x^{2^{m+1}+2}+x^2+x$;
\item[{\rm (3)}] $f(x)=x^{2^{m+2}+4}+x^{2^{m+1}+2}+x^2+x$;
\item[{\rm (4)}] $f(x)=x^{2^{2m+1}-2^{m+1}+2}+x^{2^{m+1}}+x^2+x$;
\item[{\rm (5)}] $f(x)=x^{2^{2m+1}-2}+x^{2^{2m+1}-2^{m+1}}+x^{2^{2m+1}-2^{m+1}-2}+x$.
\end{enumerate}
\end{lem}

\begin{thm}\label{thm:involution:odd}
For $m \in \N$, the following polynomials are involutions over $\bF_{2^{2m+1}}$:
\begin{enumerate}
\item[{\rm (1)}] $\cI(x)=x+\frac{1}{x^{2^{m+2}+2}+x^{2^{m+2}}+x^{2^{m+1}}+ x^2 +1}$;
\item[{\rm (2)}] $\cI(x)=x+\frac{1}{x^{2^{m+1}}+x^{2^{m+1}-1}+1}$;
\item[{\rm (3)}] $\cI(x)=x+\frac{1}{x^{2^{m+2}+2}+x^{2^{m+1}}+1}$;
\item[{\rm (4)}] $\cI(x)=x^{2^{2m+1}-2^{m+1}+1}+x^{2^{m+1}-1}+x+1$;
\item[{\rm (5)}] $$\cI(x)=\left\{
      \begin{aligned}
      &x^{2^{2m+1}-2}+x^{2^{2m+1}-2^{m+1}}+x^{2^{2m+1}-2^{m+1}-2},& & x \not \in \bF_{2}; \\
      &x+1, & & x \in \bF_{2}.
      \end{aligned}\right.$$
\end{enumerate}
\end{thm}
{\it Proof.}  We prove the conclusion only for Case (1). The conclusions in Cases (2)-(5) can be similarly proved.
Since $f(x)=x^{2^{m+1}+2}+x^{2^{m+1}}+x^2+x$ in Lemma~\ref{lem:2-1} is a \2 mapping over $\bF_{2^{2m+1}}$, by Theorem~\ref{thm:main}
we only need to check that $y=\cI(x)$ in Case (1) satisfies that $f(x)+f(y)=0$ for $x\in \bF_{2^{2m+1}}$.

Let $X=x^{2^{m+1}}$, then $X^{2^{m+1}}=x^2$ and $\cI(x)=x+\frac{1}{X^2x^2+X^2+X+x^2+1}$.
We claim that $X^2x^2+X^2+X+x^2+1\neq 0$ for $x \in \bF_{2^{2m+1}}$. Otherwise, we have
\begin{equation}\label{eq:involutionodd1} H(x):=X^2(x^2+1)+X+x^2+1= 0 \end{equation} for some $x \in \bF_{2^{2m+1}}$.
Raising the both sides of Eq.(\ref{eq:involutionodd1}) to the power of $2^{m+1}$ , we get
\begin{equation}\label{eq:involutionodd2} G(x):=X^2(x^4+1)+x^4+x^2+1= 0. \end{equation}
Combining Eq.(\ref{eq:involutionodd1}) and Eq.(\ref{eq:involutionodd2}), we obtain
$$ (x^4+1)\cdot (H(x))^2+(G(x))^2+G(x)=(x+1)^2=0.$$
This contradicts Eq.(\ref{eq:involutionodd1}). Then the conclusion follows from direct verification by substitution.
\EOP

\section{Concluding remark}\label{sec:final}

In this paper, we characterized a closed relationship between \2 mappings and involutions without any fixed point over finite fields, and proposed an AGW-like criterion for \2 mappings over finite fields of even characteristic. Using this criterion, eight classes of \2 mappings of the form~(\ref{eq:main}) were obtained, and ten classes of involutions without any fixed point were produced by the proposed connection between the \2 mappings and the involutions. Theorem~\ref{thm:main} shows that a \2 mapping can derive an involution over finite fields, but it seems not easy to obtain the explicit expression of derived involutions from \2 mappings. This is a problem worthy of further investigation.

\end{document}